\documentclass{procmult}
\usepackage{graphicx}

\begin{document}

\title{Recent developments on matter dynamics within the Kaluza-Klein picture}
\author{Valentino Lacquaniti$^\dag$,$\diamond$,$\star$ Giovanni Montani$\dag$,$^\ddag$,$\ast$}
\institute{
$\dag$ ICRA---International Center for Relativistic Astrophysics, 
Physics Department (G9),
University  of Rome, "`La Sapienza"', 
Piazzale Aldo Moro 5, 00185 Rome, Italy.\\
$\diamond$  Physics Department  "`E.Amaldi "`,  University of Rome, "`Roma Tre"', 
Via della Vasca Navale 84, I-00146, Rome , Italy \\ 
$\star$ LAPTH -9, Chemin de Bellevue BP 110 74941 Annecy Le Vieux Cedex, France \\
$\ddag$ ENEA- C. R. Frascati ( Department F. P. N. ), via E.Fermi 45, I-00044, Frascati, Rome, Italy \\
$\ast$ ICRANET - C.C. Pescara, Piazzale della Repubblica 10, I-65100, Pescara, Italy \\
}
\maketitle
\abstract{}
In this paper we propose a new approach to matter dynamics in compactified Kaluza-Klein theories. We discard the idea that the motion is geodesic and perform a simultaneous reduction of matter geometry defining the test particle via a multipole approach. In the resulting dynamics the tower of huge massive modes is removed, without giving up the compactification scenario. Such an approach yields a consistent modified gravity theory with source. Some scenarios and applications to dark energy problem are  sketched.

\section{Reduction of Matter and Geometry}
It is well known how the introduction of matter in the framework of compactified Kaluza-Klein (KK) theories \cite{kaluza},\cite{modernkk},\cite{overduinwesson} leads to an unsolved puzzle \cite{KK1},\cite{overduinwesson}. Such a problem can be summarised as follows. Let us consider the dynamics of a free 5D particle: generalizing the usual  approach that we adopt in General Relativity we describe the particle via the Action $S_{particle}=-\hat{m}\int ds_5$, being $\hat{m}$ an unknown mass parameter .  Such a picture leads to the geodesic equation $\frac{dw^{A}}{ds_5}$=0 and to the associate Klein-Gordon equation $\tilde{P}_{A}\tilde{P}^{A}=\hat{m}^2$, where $w^{A}$ and $P^{A}$ are the 5D velocity and the 5D momentum respectively. By performing the dimensional reduction of such equations we are able to achieve the dynamics of an interacting 4D particle, thus reproducing the electrodynamics, where the effective coupling factors $q$ and $m$ of the particle are defined in terms of $P_5$,  which is conserved due to the cylindricity hypothesis. In the most simple KK model, where we set as a constant the extra scalar field that appears in such theories we have:
$$
q=\sqrt{4G}P_5\quad\quad\quad  m^2=\hat{m}^2+P_5^2\,.
$$
Therefore $\hat{m}$ does not represents the physical mass of the particle.
It  is easy to see that, regardless the value of $\hat{m}$, the $q/m$ ratio is upper bounded in such a way that it cannot represents any known elementary particle. Moreover, taking into account the compactification of the extra dimension, we get a discrete spectrum for the mass, where a tower of modes proportional to the inverse of the length of the extra dimension appears. The value of the elementary charge $e$ can be restored assuming $l_5\simeq10^{-31} cm$ but at the same time such an evaluation provides massive modes beyond Planck scale, therefore there is no way to recover a consistent picture  within this framework. Indeed , various effort have been done in order to relax the original compactification assumption and nowadays non-compactified models are in full developments \cite{rubakov}. Here we propose an alternative approach which does not give up with such scenario: we still assume the compactification of the extra dimension, thus adopting the cylindricity condition as a low energy limit; in this respect we discard the idea that is possible to deal \textit {a priori} with the geodesic approach: indeed, such an approach relies on the possibility to define a test particle as a localized object \cite{pap}, but, given the compactification of the fifth dimension, such a scenario is no more guaranteed.
Our scheme is therefore the following \cite{KK2}: at first we  introduce a cylindrical, unknown, $5D$ matter tensor. Then, we  perform the dimensional reduction, in order to extract some physical meaning about these degrees of freedom and, finally, we face the rigourous definition of a particle via a multipole expansion \cite{KK1},\cite{pap}.
Hence, starting from $5D$ Einstein equation ${}^5R^{AB}=8\pi G_{5}T^{AB}$, being $G_5$ the unknown $5D$ Newton constant, we get the following set:
\begin{eqnarray}
\label{tensor}
&{}& G^{\mu\nu}=\frac{1}{\phi}\nabla^{\mu}\partial^{\nu}\phi-\frac{1}{\phi}g^{\mu\nu}g^{\alpha\beta}\nabla_{\alpha}\partial_{\beta}\phi+8\pi G\phi^2T^{\mu\nu}_{em}+8\pi G \frac{T^{\mu\nu}_{matter}}{\phi}\,, \\
\label{vector}
 &{}& 
  \nabla_{\nu}\left(\phi^3F^{\nu\mu}\right)=4\pi j^{\mu} \,,\\
  \label{scalar}
 &{}& 
  g^{\alpha\beta}\nabla_{\alpha}\partial_{\beta}\phi=- G\phi^3 F^{\mu\nu}F_{\mu\nu}+\frac83 \pi G \left( T_{matter}+2\frac{\vartheta}{\phi^2}\right)\,.
  \end{eqnarray}
In the above equations $G^{\mu\nu}$ is the usual Einstein tensor, $F^{\mu\nu}$ the Faraday tensor, $\phi$ the extra scalar field governing the expansion of the extra dimension, and, given the coordinate length of the fifth dimension $l_5=\int dx^5$, we have \cite{KK2}:
\begin{equation} 
T^{\mu\nu}_{matter}=l_5\phi T^{\mu\nu} ,\quad\quad  j^{\mu}=\sqrt{4G}l_5\phi T_5^{\mu},\quad\quad \vartheta=l_5\phi T_{55} ,\quad\quad G=G_5l_5^{-1}\,.
\end{equation}
Now, by mean of Bianchi reduced identities we also have the other set:
\begin{eqnarray}
\label{matter}
&{}&  \nabla_{\rho}( T^{\mu\rho}_{matter})=-g^{\mu\rho}\left(\frac{\partial_{\rho}\phi}{\phi^3}\right)\vartheta+ F^{\mu}_{\,\,\rho}j^{\rho}\, ,\\
\label{current}
&{}& \nabla_{\mu}j^{\mu}=0\, .
  \end{eqnarray}
 As a check of internal consistency of the model should be noticed that conservation equations (\ref{matter}, \ref{current}) can either be derived by the dimensional reduction of the $5D$ Bianchi identities or by fields equations (\ref{tensor},\ref{vector},\ref{scalar}) by mean of $4D$ Bianchi identities plus a little algebraic machinery. At this step it is worth noting that in the simplest KK scenario, \textit{i.e} $\phi=1$, the model just reproduces the Einstein-Maxwell dynamics, in presence of a conserved current $j^{\mu}$ and a matter tensor $T^{\mu\nu}_{matter}$ coupled to $F^{\mu\nu}$. Moreover, while in vacuum it is known that the condition $\phi=1$ leads to the inconsistent condition $F^{\mu\nu}F^{\mu\nu}=0$, here it simply correspond to a viable solution for the sector $\vartheta=\frac{3}{16\pi} F^{\mu\nu}F_{\mu\nu}-\frac12   T_{matter} $ \footnote{Notice that if $\phi=1$ the choice of the equation of state of the unknown matter source $\vartheta$ becomes in principle arbitrary, because in such a case eq.\ref{matter} leaves $\vartheta$ undetermined   }. In the generic case the model provide a theory of gravity modified by the presence of the scalar field $\phi$, which is usual in $5D$ theories, plus the scalar field $\vartheta$ which
 represents a new matter source term of pure extra dimensional origin.
 
 \section{Removal of the Kaluza-Klein Tower}
 To get insights on the physical meaning of the extra degrees of freedom, and to check the consistency of the model we need to face the problem of particle motion. Such a task can be accomplished starting from eq. \ref{matter},\ref{current} and performing a multipole expansion \textit{a l\'a} Papapetrou \cite{pap}, thus describing the motion of the test particle by mean of the single-pole equation. Assuming that $T^{\mu\nu}$ and $j^{\mu}$ are localized on a 4D trajectory $X^{\mu}$ the resulting equation reads \cite{KK1}
 \begin{equation}
m\frac{Du^{\mu}}{Ds}=A(u^{\rho}u^{\mu}-g^{\mu\rho})\frac{\partial_{\rho}\phi}{\phi^3}+qF^{\mu\rho}u_{\rho}\,,
\label{neweq}
\end{equation}
where $u^{\mu}=\frac{dX^{\mu}}{ds}$, $ds^2=g^{\mu\nu}dX^{\mu}dX^{\nu}$.  The coupling factors $m$, $q$, $A$ and the effective tensor describing the test particles read as follows:\\
{\small
\begin{minipage}[c]{.45\textwidth}
\begin{eqnarray*}
 m&\!=\!&\frac{1}{u^0}\int\!\!\!d^3x\,\sqrt{g}\, T^{00}_{matter} \\
q&\!=\!&\int\!\!\!d^3x\,\sqrt{g}\,j^{0} \\
 A&\!=\!&u^0\int\!\!\!d^3x\,\sqrt{g}\, \vartheta
\end{eqnarray*}
\end{minipage}\begin{minipage}[c]{.45\textwidth}
\begin{eqnarray*}
  \sqrt{g} T^{\mu\nu}_{matter}&\!=\!&\int\!\!\!ds\,m\,\delta^4(x-X)u^{\mu}u^{\nu} \\
  \sqrt{g} j^{\mu}&\!=\!&\int\!\!\!ds\,q\,\delta^4(x-X)u^{\mu} \\
  \sqrt{g}\vartheta&\!=\!&\int\!\!\!ds\,A\,\delta^4(x-X)
\end{eqnarray*}
\end{minipage}}\\
Hence, being the components of the matter tensor localized just in the ordinary 4D space the particles turns out to be delocalized into the fifth dimension. The $q/m$ ratio is no more bounded because now $q$ and $m$ are defined in terms of independent degrees of freedom while within the geodesic procedure they were both depending on $P_5$. Charge is still conserved while mass, as expected in a theory with a scalar-tensor coupling, is now variable and its behaviour is governed by the equation
\begin{equation}
\frac{dm}{ds}=-\frac{A}{\phi^3}\frac{d\phi}{ds} \, .
\label{varmass}
\end{equation}
The new equation (\ref{neweq}) admits an effective Action which reads:
\begin{equation}
S_{particle}=-\int\!\!m\,ds+q(A_{\mu}dx^{\mu}+\frac{dx^5}{\sqrt{4G}}).
\end{equation}
In consequence of eq. \ref{varmass} now $m$ is no more constant and this take into account the fact that in KK model the 5D Equivalence Principle is broken. Starting from the above Action we can calculate conjugate momenta and dispersion relation which are in $5D$ and $4D$ formalism respectively:\\
\begin{minipage}[c]{.45\textwidth}
$$
P_AP^{A}=m^2-\frac{q^2}{4G\phi^2} \quad\quad\quad\quad \Longleftrightarrow
$$ 
\end{minipage}\begin{minipage}[c]{.45\textwidth}
\begin{equation}
\left\{
\begin{array}{l}
\Pi_{\mu}\Pi^{\mu}=m^2\\
\\
P_5=\frac{q}{\sqrt{4G}}\\
\\
\Pi_{\mu}=P_{\mu}-qA_{\mu}
\end{array}
\right.
\end{equation}
\end{minipage}\\
Looking at $4D$ relations we recognize the minimal substitution typical of electrodynamics and we see that $m$ represents the physical mass of the particle. Moreover, taking now into account the compactification of the extra dimension, we can still have an explanation for the discretization of the charge, with an evaluation of $10^{-31} cm$ for the size of the fifth dimension, but this does not affects the mass which is not dependent on $P_5$. At the same time, looking at the 5D dispersion relation we recognize the presence of an additional term with respect to the mass term.
By studying the associated Klein-Gordon dynamics, via canonical quantization, we see that such an extra contribute acts as a counter-term and rules out the tower of massive modes. 
Hence, within this revised approach the KK tower is removed and it becomes possible to deal with matter without renouncing to the compactification hypothesis. The production of the tower is an unphysical result provided by the geodesic procedure which is a misdealing approach because, relying on the definition of a $5D$ localized particle, it
contradicts the starting hypothesis of compactification.

\section{Physical Remarks}
Given the possibility to deal consistently with matter let us turn back to the starting equations \ref{tensor}, \ref{vector}, \ref{scalar}, and \ref{matter}, \ref{current}. As noted above, we now deal with a modified gravity theory \cite{copeland}, where it seems interesting the possibility to address the metric field $\phi$ to dark energy and the matter field $\vartheta$ to dark matter. Here we focus just on some simple scenarios in absence of electromagnetic fields: equations of  particular interest are \ref{scalar}, \ref{neweq}, \ref{varmass} which in such a case read:
\begin{equation}
g^{\alpha\beta}\nabla_{\alpha}\partial_{\beta}\phi=\frac83 \pi G \left( T_{matter}+2\frac{\vartheta}{\phi^2}\right)\, ,
\end{equation}
\begin{equation}
\frac{dm}{ds}=-\frac{A}{\phi^3}\frac{d\phi}{ds}\,, \quad\quad m\frac{Du^{\mu}}{Ds}=A(u^{\rho}u^{\mu}-g^{\mu\rho})\frac{\partial_{\rho}\phi}{\phi^3}\, .
\end{equation}
Interesting equations of state we would like to outline are the following:
\begin{itemize}
\item $2\vartheta=-\phi^2 T_{matter}$ \\
 Here $\phi=1$ is a suitable solution and it yields $m=cost$, $\frac{Du^{\mu}}{Ds}=0$; therefore the Free Falling Universality ( FFU ) of particles still holds and we just  recover  General Relativity. 

\item {$\vartheta=0 $} \\
Now $m=cost$, being $A=0$ and we have $\frac{Du^{\mu}}{Ds}=0$. Therefore the FFU holds, but we can have $\phi$ variable, thus we have a modified theory.

\item {$A=\alpha m\phi^2$} \\
Now $\phi$ is variable as well as $m$  but the equation of motion is: $\frac{Du^{\mu}}{Ds}=\alpha(u^{\rho}u^{\mu}-g^{\mu\rho})\frac{\partial_{\rho}\phi}{\phi}$. The mass is ruled out and then FFU still holds even if the theory is now modified by two additional degrees of freedom.
\end{itemize}
Noticeably, in the last scenario the equation for mass behaviour admits an easy integration and we get  a scaling law for mass:
\begin{equation}
m=m_0\left(\frac{\phi}{\phi_0}\right)^{-\alpha}\, .
\label{scalmass}
\end{equation}
Therefore it looks like that we can manipulate the degree of modification of the theory by addressing various kind of state equations to the unknown source $\vartheta$; in most interesting cases however we still have a constant mass or, at least the FFU.
A natural development of such a study , in order to probe the model and check its relation to dark matter and energy models, is to implement the dynamics into simplified backgrounds like the homogeneous one or the spherically symmetric one. First studies in such a direction are encouraging.
Indeed, it is possible to show that, as far as the homogeneous scenario is concerned, the comoving velocity $U^{\mu}= (1,0,0,0)$ is still a solution of the motion equation,  regardless of the presence of scalar fields into the motion equations \ref{neweq}. Such a result allows us to still adopt the definition $T^{\mu\nu}_{matter}=(\rho+p)U^{\mu}U^{\nu}-g^{\mu\nu}p$ for a perfect fluid described in terms of its pressure and density, and suggests to consider a simple parametrization of the extra source $\vartheta$ like $\vartheta=\phi^2\left(\alpha \rho +\beta p\right)$. Such a parametrization mimics the equation of state $p=\gamma\rho$ and at the same time it yields, for a dust matter,   the scenario $A=\alpha m \phi^2$ we discussed above.
On the other side, looking at the exterior solution in a spherically symmetric background ( Generalized Schwarzschild Solution, \cite{sorkin} ) it is worth remarking that the behaviour of the mass distribution is given in the exterior region by an equation of the form
 $m(r)\, \infty  \,\phi^{-1}$,
 which thus coincides to our equation  \ref{scalmass}  for $\alpha=1$.
 
 Hence, in conclusion, we think that this new approach offers a viable framework to deal consistently with matter in the scheme of compactified models. When the case $\phi=1$ is considered the theory provides a toy-model for a more general unification scheme, while, in the general case it provides a consistent modified gravity theory which is able to not broke the FFU of particles. The study of the model is in progress and in our opinion it deserves future effort to be pursued.


\begin{thebibliography}{99}


\bibitem{kaluza}
T. Kaluza, \emph{On the Unity Problem of Physics}, Sitz. Press. Akad. Wiss. Phys. Math. ,1921.\\
O.Klein, \emph{Z.F.Physik}, \textbf {37},  1926.\\
O.Klein, \emph{Nature }, \textbf{118}, 1926.

\bibitem{modernkk}
T.Appelquist, A.Chodos, P.Frund, \emph{Modern Kaluza-Klein theories}. Addison Wesley, 1987.

 \bibitem{overduinwesson}
J.Overduin, P. Wesson,  {\it Phys. Rep.},  {\bf 283}, (1997), 303. arXiv:gr-qc/9805018 v1.

\bibitem{KK1}
 V. Lacquaniti, G. Montani, \emph{Dynamics of Matter in a Compactified 5D Kaluza-Klein Model }, in press, \emph{Int. J. Mod .Phys. D}, 2009. ArXiv:0902.1718v1 [gr-qc].

\bibitem{rubakov}
V. A. Rubakov,
	\emph{Phys.Usp.},\textbf {44}, (2001) 871-893.	arXiv:hep-ph/0104152v2.
	
\bibitem{pap}
A. Papapetrou, \emph{ Proc. Phys. Soc. } \textbf{64}, 57 (1951).



\bibitem{KK2}
 V. Lacquaniti, G. Montani, \emph{Geometry and Matter Reduction in a 5D Kaluza-Klein Framework},  to appear on \emph{Int. J. Mod .Phys. A}, 2009. ArXiv:0906.0804v1 [gr-qc]










\bibitem{copeland}
E.J. Copeland, M.Sami, S.Tsujikawa,\emph{Int.J.Mod.Phys.D},\textbf{15},1753,1936, (2006).





\bibitem{sorkin}
R. D. Sorkin,  \emph{Phys. Rev. Lett.}, \textbf{ 51}, (1983) 87. \\
D. J. Gross , M. J. Perry, \emph{Nucl. Phys.}, \textbf{ B226}, (1983) 29. \\
A. Davidson, D. A. Owen, \emph{Phys. Lett.}, \textbf{155B}, (1985) 247. \\
P. S. Wesson , J. Ponce de Leon, \emph{ Class. Quant. Grav.}, \textbf{ 11}, (1994) 1341. \\
P. S. Wesson, \emph{ Astrophys. 
J.}, \textbf{ 420}, (1994) L49. 




















\end{thebibliography}
\end{document}